\documentclass[aps,prc,preprint,superscriptaddress,showpacs]{revtex4}
\usepackage{amsmath}
\usepackage{amssymb}
\usepackage{times}
\usepackage{mathrsfs}
\usepackage{multirow}
\usepackage{bm}
\usepackage{ulem}
\usepackage{color}
\usepackage[dvipdfmx]{graphicx}

\def\la{\mathrel{\mathpalette\fun <}}
\def\ga{\mathrel{\mathpalette\fun >}}
\def\fun#1#2{\lower3.6pt\vbox{\baselineskip0pt\lineskip.9pt
\ialign{$\mathsurround=0pt#1\hfil##\hfil$\crcr#2\crcr\sim\crcr}}}

\def\a{\alpha}

\newcommand{\beq}{\begin{equation}}
\newcommand{\eeq}{\end{equation}}
\newcommand{\bea}{\begin{eqnarray}}
\newcommand{\eea}{\end{eqnarray}}

\newcommand{\bfi}[1]{\mbox{\boldmath $#1$}}
\newcommand{\bfis}[1]{\mbox{\boldmath ${\scriptstyle #1}$}}

\newcommand{\vk}{{\bfi k}}

\newcommand{\vs}{{\bfi s}}

\newcommand{\vrr}{{\bfi r}}
\newcommand{\vR}{{\bfi R}}

\newcommand{\vik}{{\bfis k}}

\newcommand{\viR}{{\bfis R}}

\def\bk{\mbox{\boldmath $k$}}

\begin{document}

\title{
Roles of chiral three-nucleon forces in nucleon-nucleus scattering
}

\author{Masakazu Toyokawa}
\email[]{toyokawa@phys.kyushu-u.ac.jp}
\affiliation{Department of Physics, Kyushu University, Fukuoka 812-8581, Japan}

\author{Kosho Minomo}
%\email[]{minomo@rcnp.osaka-u.ac.jp}
\affiliation{Research Center for Nuclear Physics (RCNP), Osaka
University, Ibaraki 567-0047, Japan}

\author{Michio Kohno}
%\email[]{kohno@kyu-dent.ac.jp}
\affiliation{Research Center for Nuclear Physics (RCNP), Osaka
University, Ibaraki 567-0047, Japan} 
\affiliation{Physics Division, Kyushu Dental University, Kitakyushu 803-8580, Japan}

\author{Masanobu Yahiro}
%\email[]{yahiro@phys.kyushu-u.ac.jp}
\affiliation{Department of Physics, Kyushu University, Fukuoka 812-8581, Japan}

\date{\today}

\begin{abstract} 
We investigate the effects of chiral three-nucleon force (3NF) at NNLO level 
on nucleon-nucleus (NA) elastic scattering, 
using the standard framework based on 
the Brueckner-Hartree-Fock method for nuclear matter and 
the $g$-matrix folding model for NA elastic scattering. 
The optical potential in nuclear matter calculated 
from chiral two-nucleon force (2NF) at N$^{3}$LO level is found to be 
close to that from Bonn-B 2NF, whereas the Melbourne $g$-matrix is known as 
a practical effective nucleon-nucleon interaction constructed by localizing the $g$-matrices calculated from Bonn-B 2NF. 
As the first attempt to estimate chiral-3NF effects on NA scattering, 
the effects are simply introduced by multiplying the local Melbourne $g$-matrix 
by the ratio of the optical potential in nuclear matter calculated 
from chiral 2NF+3NF to that from chiral 2NF. 
For NA elastic scattering on various targets at 65 MeV, 
chiral 3NF  
makes the folding potential less attractive and more absorptive.
The novel property for the imaginary part is 
originated in the enhancement of tensor correlations 
due to chiral 3NF (mainly the 2$\pi$-exchange diagram). 
The two effects are small for differential cross sections and 
vector analyzing powers at the forward and middle angles 
where the experimental data are available. 
If backward  measurements are made, the data will reveal the effects 
of chiral 3NF. 
\end{abstract}

%21.30.Fe  Forces in hadronic systems and effective interactions
%24.10.Ht  Optical and diffraction models
%25.70.Bc  Elastic and quasielastic scattering
%25.40.Cm  Elastic proton scattering
%25.40.Dn  Elastic neutron scattering

\maketitle

%Introduction
\section{Introduction}
\label{Introduction}

Microscopic understanding of nucleon-nucleus (NA) and nucleus-nucleus (AA) 
optical potentials is one of the primary goals of nuclear physics. 
The optical potentials play an important role not only 
in describing elastic scattering but also in analyzing 
more complicated reactions, since the potentials are inputs 
of theoretical calculations such as the distorted-wave Born approximation
and the continuum discretized coupled-channels method
\cite{CDCC-review1,CDCC-review2,Yahiro:2012tk} 
for inelastic scattering and transfer and breakup reactions.

NA elastic scattering is less absorptive and hence more transparent 
than AA elastic scattering. In this sense, 
NA elastic scattering is more informative. 
From a theoretical viewpoint based on the multiple scattering 
theory~\cite{Watson,KMT,Yahiro-Glauber}, furthermore, 
the multiple nucleon-nucleon (NN) collision series 
 is much simpler in NA scattering~\cite{Watson,KMT} than in 
AA scattering~\cite{Yahiro-Glauber}. Microscopic understanding is thus easier 
for NA scattering than for AA scattering. 
In this paper, we focus our discussion on NA scattering.

The $g$-matrix folding model is a useful tool of describing 
NA scattering~\cite{M3Y,JLM,Brieva-Rook,Satchler-1979,Satchler,CEG,
Rikus-von Geramb,Amos,CEG07}. 
In the model, the optical potential is obtained by folding 
the $g$-matrix interaction with the target density. 
Since the $g$-matrix is evaluated 
in nuclear matter, the local-density approximation is taken 
in the folding procedure. Target-excitation and Pauli-blocking effects are 
included within the approximation. 
Among various kinds of $g$-matrix interactions, the Melbourne 
$g$-matrix is successful in reproducing the experimental data 
on cross sections and spin observables systematically 
without introducing any ad hoc phenomenological adjustment~\cite{Amos}. 
This is a monumental achievement in nuclear reaction studies.

The microscopic optical potential calculated with the $g$-matrix folding model 
is nonlocal and hence not practical in many applications, 
but it can be localized with the Brieva-Rook approximation~\cite{Brieva-Rook}. 
The validity of the approximation is shown in a wide range of 
incident energies~\cite{Minomo:2009ds}.
The local version of the Melbourne $g$-matrix folding potential is consistent 
with the phenomenological optical potentials~\cite{Toy13}.

Another important issue in nuclear physics is to clarify the roles of 
three-nucleon force (3NF) in finite nuclei, nuclear reactions and 
nuclear matter. 
The phenomenological approach to this issue began with 
the 2$\pi$-exchange 3NF proposed by Fujita and Miyazawa~\cite{Fuj57}. 
Attractive 3NFs were introduced to reproduce the binding 
energies for light nuclei ~\cite{Wir02}, whereas 
repulsive 3NFs were used to explain the empirical saturation properties 
in symmetric nuclear matter~\cite{Wir88,Mut00,Dew03,Bog05}. 
Recently, chiral effective field theory (Ch-EFT) made a theoretical 
breakthrough in this issue~\cite{Epelbaum-review-2009,Machleidt-2011}. 
The theory provides a systematic low momentum 
expansion based on chiral perturbation theory to interactions among nucleons. 
This allows us to define two-nucleon force (2NF) and 3NF definitely. 
The roles of chiral 3NF are investigated, for example, 
in Refs~\cite{Epe02,Nog06,Nav07-FBS,Nav07,Ski11,Kalantar-2012} 
for light nuclei 
and in Refs.~\cite{Sam12,Koh12,Car13,Koh13} for nuclear matter. 
The $g$-matrix calculated with 
the Brueckner-Hartree-Fock (BHF) method from chiral 2NF+3NF 
is successful in reproducing the empirical equation of state (EoS) 
of symmetric nuclear matter~\cite{Sam12,Koh12,Car13,Koh13}. 
In the framework, the effects of chiral 3NF appear 
through density ($\rho$) dependence of the $g$-matrix.

The effects of 3NF on NA elastic scattering were investigated with 
CEG07 $g$-matrix \cite{CEG07} that is constructed from 2NF based on 
the Nijmegen extended soft-core model~\cite{Rij06a,Rij06b}. 
The effects of 3NF  
are effectively taken into account by introducing the $\rho$-dependent 
vector-meson mass that reproduces the empirical EoS. 
Very recently, the effects of 3NF on NA scattering were 
investigated~\cite{Raf13} 
with the $g$-matrix calculated from AV18 2NF~\cite{Wir95} plus 
phenomenological 3NFs~\cite{Pud97,Fri81,Lag81}. 
The 3NF improves the agreement with measured vector analyzing powers.
In these approaches, however, 
the real and/or imaginary parts of the folding 
potential are adjusted to measured cross sections.

In this paper, we investigate the roles of chiral 3NF in 
NA elastic scattering. 
For this purpose, nuclear matter 
calculations are done for positive energy ($E$) 
by using chiral N$^{3}$LO  2NF including NNLO 3NF 
with the cutoff of 550 MeV~\cite{Epe05,HEB11} that well reproduces 
empirical saturation properties of symmetric nuclear matter 
for negative $E$~\cite{Koh12,Koh13}. 
Even for 2NF, it is quite difficult to treat three-nucleon correlations 
in nuclear matter calculations. 
Hence we make the mean-field approximation, that is, we derive 
an effective 2NF from 3NF by averaging it over 
the third nucleon in the Fermi sea. 
The approximation is considered to be good for nucleon elastic scattering 
in nuclear matter, since nucleons in the Fermi sea are not excited 
in the final stage of the scattering. 
The optical (single-particle) potential for the scattering 
is then calculated from the sum of original and reduced 2NFs 
by using the BHF method.

The effects of chiral 3NF can be described by the ratio $f$ of 
the single-particle potential ${\cal U_{\rm (2NF+3NF)}}$ calculated 
from chiral 2NF+3NF to the potential ${\cal U_{\rm (2NF)}}$ 
from chiral 2NF. The single-particle potential is nothing but 
the optical potential in nuclear matter. 
The potential ${\cal U_{\rm (2NF)}}$ is found to be close to
the single-particle potential calculated from Bonn-B 2NF \cite{BonnB}, 
whereas the Melbourne $g$-matrix is known as a local 
effective NN interaction obtained 
by localizing the $g$-matrices calculated 
from Bonn-B 2NF \cite{Amos}.  
We then simply incorporate the chiral-3NF effects 
in the local Melbourne $g$-matrix 
by multiplying the $g$-matrix by the factor $f$,  
as the first attempt to estimate the effects on NA scattering.  
The chiral-3NF effects are investigated 
at a lower incident energy of $E=65$ MeV over various targets, 
since Ch-EFT is more appropriate for lower $E$.

In Sec. \ref{Theoretical framework}, we recapitulate the BHF method 
for the symmetric nuclear matter with both 2NF and 3NF 
and the $g$-matrix folding model for the NA system. 
The modified Melbourne $g$-matrix with chiral-3NF corrections is also 
presented. In Sec. \ref{Results}, the results of the $g$-matrix folding model 
are shown for NA scattering. Section \ref{Summary} is devoted to a summary.

%Theoretical framework
\section{Theoretical framework}
\label{Theoretical framework}

\subsection{$g$-matrix calculations for 3NF}
\label{$g$-matrix calculations for 3NF}
We recapitulate the BHF method for nuclear matter with 2NF plus 3NF, 
following Refs. \cite{Koh12,Koh13}. 
It is quite difficult to treat 3NF $V_{123}$ in infinite matter. 
We then derive an effective 2NF $V_{12(3)}$  from chiral 3NF, using 
the mean-field approximation \cite{Fri81,KAT74,LNR71,HKW10}, 
that is, $V_{123}$ is 
averaged over the third nucleon in the Fermi sea: 
\bea
 \langle \bk_1' ,\bk_2' |V_{12(3)}| \bk_1 , \bk_2 \rangle_A 
 \equiv \sum_{\vik_3} \langle \bk_1' , 
 \bk_2' , \bk_3 |V_{123}| \bk_1 ,  \bk_2 , \bk_3 \rangle_A, 
\eea
where the suffix $A$ denotes the antisymmetrization and the symbol 
$\bk_i$ stands for quantum numbers (momentum and $z$ components of spin and 
isospin) of the $i$th nucleon. 

The potential energy is evaluated as 
\begin{eqnarray}
&&\frac{1}{2} \sum_{\vik_1 \vik_2} \langle \bk_1 \bk_2 | V_{12} |\bk_1 \bk_2\rangle_A 
+ \frac{1}{3!}\sum_{\vik_1 \vik_2 \vik_3} \langle \bk_1 \bk_2 \bk_3| V_{123} |\bk_1 \bk_2 \bk_3\rangle_A
 \nonumber \\
&&= \frac{1}{2} \sum_{\vik_1 \vik_2} \langle \bk_1 \bk_2 | V_{12}+\frac{1}{3} V_{12(3)} |\bk_1 \bk_2\rangle_A .
\label{pot-energy}
\end{eqnarray}
This means that 
the $g$-matrix $g_{12}$ should be calculated by 
\begin{equation}
 g_{12}=V_{12}^{\rm eff}+V_{12}^{\rm eff} G_0 g_{12} 
\label{g-eq}
\end{equation}
with the effective 2NF
\bea
V_{12}^{\rm eff}=V_{12}+ \frac{1}{3}V_{12(3)} 
\eea 
and the nucleon propagator 
\bea
G_0=\frac{Q}{E - H}, 
\eea
where $Q$ stands for the Pauli exclusion operator. 
The $g$-matrix equation \eqref{g-eq} is solved by taking 
the following continuous prescription for intermediate states. 
In Eq. \eqref{g-eq} for the component $g_{12}|\bk_1 \bk_2\rangle$, 
the nucleon propagator 
is described for intermediate nucleons 
with momenta $\bk_1'$ and $\bk_2'$ as 
\bea
\frac{1}{E-H}|\bk_1' \bk_2' \rangle = \frac{1}
{e_{\vik_1}+e_{\vik_2}-e_{\vik_1'}-e_{\vik_2'}} .
\eea
Here the single-particle energy $e_{\bk}$ 
for nucleon with momentum ${\bk}$ is defined by 
\bea
e_{\vik}=\langle \bk |t|\bk\rangle + {\rm Re}[{\cal U}(\bk)], 
\eea
where $t$ denotes the kinetic-energy operator of nucleon and 
${\cal U}$ stands for the single-particle potential defined by~\cite{Koh13} 
\bea
{\cal U}(\bk) = \sum_{\vik'}^{k_{\rm F}^{}} \langle \bk \bk' | g_{12} 
+\frac{1}{6} V_{12(3)}(1+G_0 g_{12}) 
|\bk \bk' \rangle_A .
\label{single-particle-pot-0}
\eea
The single-particle energy is nothing but the optical potential 
in nuclear matter. 
Holt {\it et al.} evaluated the single-particle potential ${\cal U}$ 
within the framework of Ch-EFT by using 
the second-order perturbation~\cite{Hol13}. 
Our formulation is consistent with theirs. The factor $1/6$ in Eq. 
\eqref{single-particle-pot-0} is important for the consistency 
\cite{Koh12,Koh13}.

In actual calculations, the partial wave expansion \cite{HT70} is taken with
the angle-average approximation to $Q$. The validity is shown 
in Ref.~\cite{SOKN00}. 
Partial waves up to the total angular momentum $J=7$ and 
the orbital angular momentum $\ell=7$ are taken into account 
in the calculations. 
The low-energy constants of chiral interactions are taken from those 
of the J\"{u}lich group \cite{Epe05}:
$c_1=-0.81$ GeV$^{-1}$, $c_3=-3.4$ GeV$^{-1}$,
and $c_4=3.4$ GeV$^{-1}$. The other constants are taken from Ref. \cite{HEB11}:
$c_D=-4.381$ and $c_E=-1.126$. The original and effective 2NFs, $V_{12}$ and 
$V_{12(3)}$, are regularized with the common form factor 
$\exp\{-(q'/\Lambda)^6-(q/\Lambda)^6\}$ with $\Lambda=550$~MeV.

There are several different sets of low-energy constants in the 
literature \cite{Nav07-FBS,Entem-2003}. 
The values of these constants are essentially same. 
In addition, as was shown in Ref. \cite{Koh13}, the variation of
$G$-matrices in nuclear matter in the 2NF level is much reduced 
when the 3NF effects are consistently incorporated. 
As for the parameters $c_D$ and $c_E$, the net effect of these terms is small, 
as far as the relation $c_D \simeq 4c_D$ holds. This relation is 
actually realized in various calculations 
in light nuclei \cite{Nav07,Noga-2006} and in nuclear matter \cite{HEB11}. 
The results of BHF calculations do not depend on specific values 
of $c_D$ and $c_E$, as far as $c_D \simeq 4c_D$ is held.

The $g$-matrix thus obtained is a function 
of the starting energy $E$ and the Fermi momentum $k_{\rm F}$, 
and can be classified as $g^{ST}(k_{\rm F},E)$ 
with the total-spin ($S$) and total-isospin ($T$) of the NN system. 
We can then decompose the single-particle potential ${\cal U}$ into 
\bea
{\cal U}=\sum_{ST}(2S+1)(2T+1){\cal U}^{ST}, 
\eea
with ${\cal U}^{ST}$ obtained from $g^{ST}$ as
\bea
{\cal U}^{ST}(k_{\rm F}^{},E)=\sum_{\vik'}^{k_{\rm F}^{}}
\langle \vk \vk'| g^{ST} 
{} + \frac{1}{6} V_{12(3)}^{ST}(1+G_0 g^{ST}) |\vk\vk'\rangle_{A} ,
\label{spp}
\eea
where $\vk$ is related to $E$ as 
$E=(\hbar \vk)^2/(2m) + {\rm Re}[{\cal U}]$ for the nucleon mass $m$. 
Here ${\cal U}^{ST}$ represents the single-particle potential 
in each spin-isospin channel. 
For the symmetric nuclear matter where 
the proton density $\rho_{p}$ is the same as the neutron one $\rho_{n}$, 
the Fermi momentum $k_{\rm F}$ is 
related to the matter density $\rho=\rho_{p}+\rho_{n}$ 
as $k_{\rm F}^3= 3\pi^2 \rho/2$ 
and hence the normal density $\rho=\rho_0=0.17$ fm$^{-3}$ corresponds to 
$k_{\rm F}=1.35$ fm$^{-1}$. 
On the right hand side of Eq. \eqref{spp}, 
the second term is ten times as small as the first term at the normal density. 
In the first term as the main component, the potential ${\cal U}^{ST}$ is 
determined 
from the on-shell component of $g^{ST}$.

%----------------------
% Figure single particle potential
%----------------------
\begin{figure}[tbp]
\begin{center}
 \includegraphics[width=0.50\textwidth]{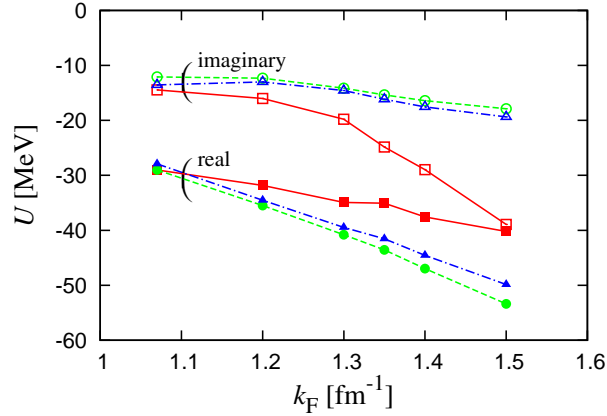}
 \caption{(Color online)
$k_{\rm F}$ dependence of the single-particle potentials ${\cal U}$ 
at $E=65$ MeV. 
The squares, circles, and triangles show the results of chiral 2NF+3NF, 
chiral 2NF and Bonn-B 2NF, respectively.
The closed (open) symbols correspond to the real (imaginary) part 
of single-particle potential.
}
 \label{fig-SPP-tot}
\end{center}
\end{figure}
%----------------------

%----------------------
% Figure Ch-3NF diagram
%----------------------
\begin{figure}[tbp]
\begin{center}
 \includegraphics[width=0.48\textwidth,clip]{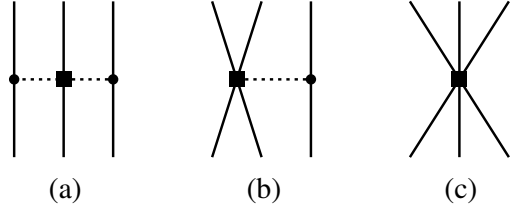}
 \caption{ Diagrams for NNLO 3NF. 
The solid and dashed lines are nucleon and pion propagations, 
respectively. Diagrams (a), (b), and (c) correspond to the 2$\pi$-exchange, 
 the 1$\pi$-exchange, and the contact interactions.
}
 \label{fig:diagram}
\end{center}
\end{figure}
%----------------------

Figure \ref{fig-SPP-tot} shows $k_{\rm F}$ dependence of the single-particle 
potential ${\cal U}$ at $E=65$ MeV. 
The squares, circles, and triangles denote the results of chiral 2NF+3NF, 
chiral 2NF and Bonn-B 2NF, respectively. 
The effects of chiral 3NF are shown by the difference between 
squares and circles. 
The difference mainly comes from the 2$\pi$-exchange 3NF 
shown by diagram (a) in Fig. \ref{fig:diagram}. 
The effects become significant 
in the region $k_{\rm F} \ga 1.2$ fm$^{-1}$. 
Particularly for NA scattering,  the chiral-3NF effects in the region 
$1.2 \la k_{\rm F} \la 1.35$ fm$^{-1}$ ($0.7 \rho_0 \la \rho \la \rho_0$) 
affect the scattering. 
In addition, the result of Bonn-B 2NF (triangles) agrees well 
with that of chiral 2NF 
(circles) at $k_{\rm F} < 1.35$ fm$^{-1}$ important for the NA scattering.

%----------------------
% Figure single particle potential
%----------------------
\begin{figure}[tbp]
\begin{center}
 \includegraphics[width=0.48\textwidth]{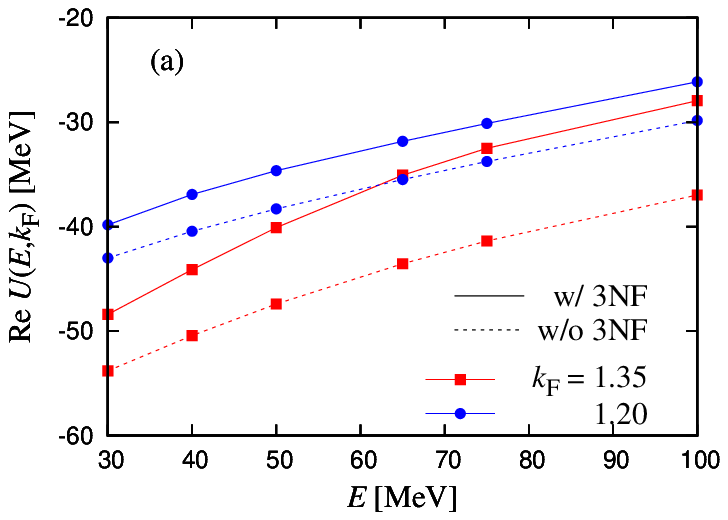}
 \includegraphics[width=0.48\textwidth]{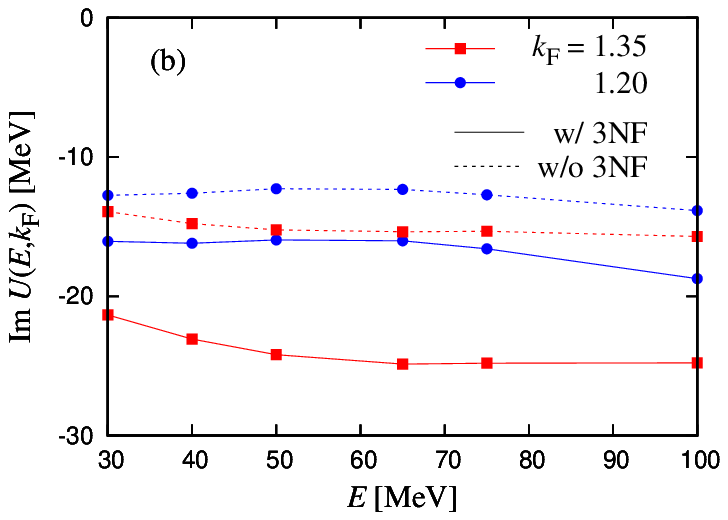}
 \caption{(Color online)
$E$ dependence of the single-particle potentials ${\cal U}$ 
at $k_{\rm F}=1.35$~fm$^{-1}$ ($\rho=\rho_0$) and 
$k_{\rm F}=1.2$~fm$^{-1}$ ($\rho=0.7\rho_0$). 
Panels (a) and (b) represent the real and imaginary strength, 
respectively. 
The results of chiral 2NF+3NF (chiral 2NF) are denoted by 
the solid (dashed) lines with symbols. 
The square (circle) symbols correspond to the results of 
$k_{\rm F}=1.35$~fm$^{-1}$ ($1.2$~fm$^{-1}$). 
}
 \label{fig-SPP-E-dep}
\end{center}
\end{figure}
%----------------------

Figure \ref{fig-SPP-E-dep} shows ${\cal U}$ as a function of $E$ for 
$k_{\rm F}=1.35$~fm$^{-1}$ ($\rho=\rho_0$) and 
$k_{\rm F}=1.2$~fm$^{-1}$ ($\rho=0.7\rho_0$). 
At low $E$ such as $0 < E \la 20$ MeV, 
the $g$-matrix in nuclear matter may not describe in-medium effects 
in finite nuclei accurately, 
since the energy levels are discrete in finite nuclei but continuous 
in nuclear matter. We then take the energy range $30 <E<100$~MeV in 
Fig. \ref{fig-SPP-E-dep}. Chiral-3NF effects are shown by the difference 
between the solid line and the corresponding dashed line. The effects are 
significant in the energy range for both cases of $k_{\rm F}=1.35$ 
and 1.2~fm$^{-1}$, although the effects become small as $k_{\rm F}$ decreases.

The single-particle potential within the framework of Ch-EFT is also 
presented by Holt {\it et al}.~\cite{Hol13} in the second-order perturbation.
It is interesting to compare our single-particle potentials with those of 
Holt \textit{et al.} (Figs. 8 and 9 of Ref.~\cite{Hol13}). In the case 
of $\rho=\rho_0$, their potential is similar to ours for the real strength. 
For the imaginary strength, meanwhile, our case is more absorptive than that of Holt \textit{et al.} in the range $30 <E<100$~MeV. 
For example, it differs by a factor of about 2 at $E=65$ MeV.
This may be due to the full ladder-summation in our g-matrix calculation.

%----------------------
% Figure single particle potential
%----------------------
\begin{figure}[tbp]
\begin{center}
 \includegraphics[width=0.45\textwidth]{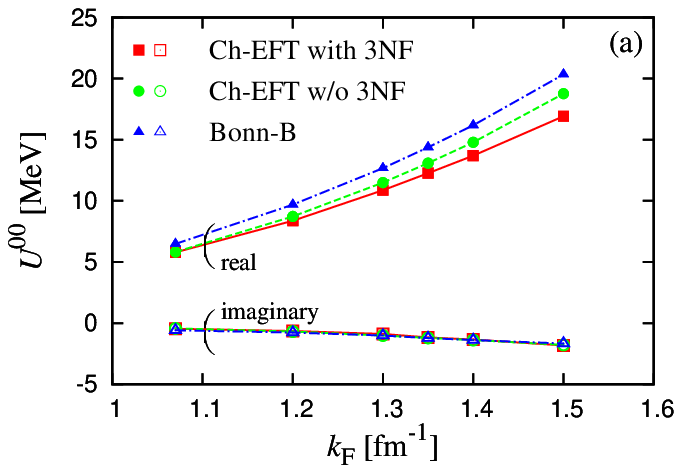}
 \includegraphics[width=0.45\textwidth]{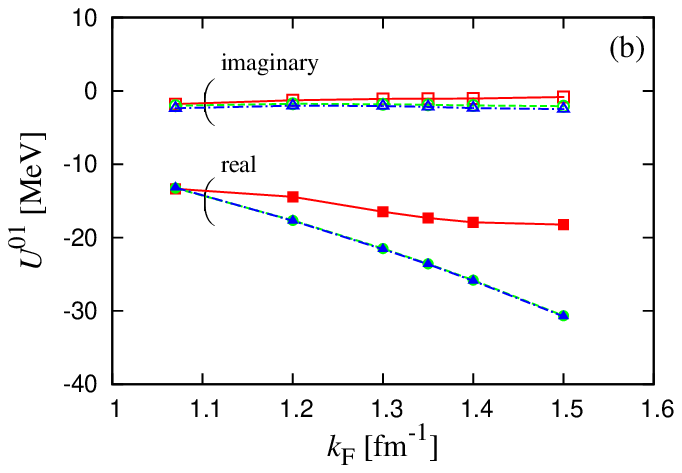}
 \includegraphics[width=0.45\textwidth]{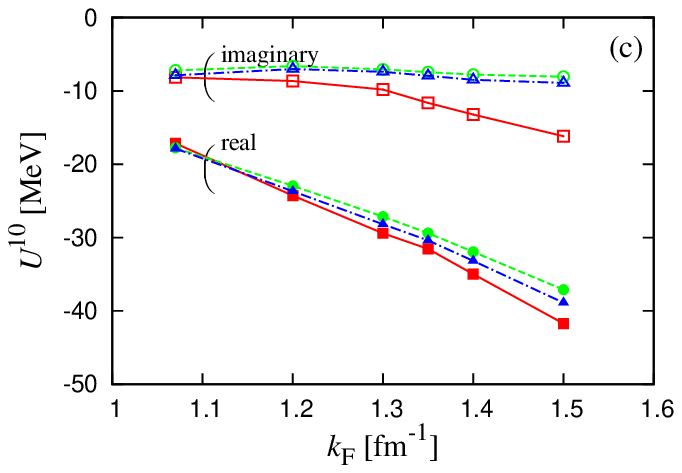}
 \includegraphics[width=0.45\textwidth]{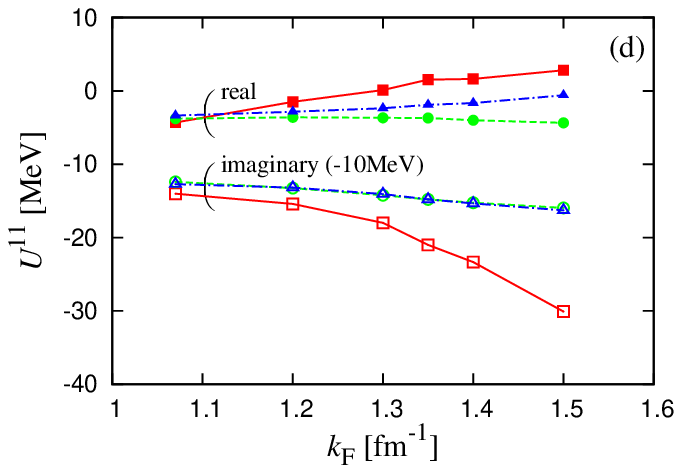}
 \caption{(Color online)
$k_{\rm F}$ dependence of ${\cal U}^{ST}$ 
at $E=65$ MeV  
for (a) $^{1}$O, (b) $^{1}$E, (c) $^{3}$E, and (d) $^{3}$O channel. 
The squares, circles, and triangles show the results of chiral 2NF+3NF, 
chiral 2NF and Bonn-B 2NF, respectively.
The closed (open) symbols correspond to the real (imaginary) part 
of single-particle potential.
For $^{3}$O, the imaginary part is shifted down by 10 MeV. 
}
 \label{fig-SPP}
\end{center}
\end{figure}
%----------------------

Now we decompose ${\cal U}$ into the ${\cal U}^{ST}$ to see the detail of 
chiral-3NF effects in Fig. \ref{fig-SPP}. Again, 
the squares, circles, and triangles correspond to 
the results of chiral 2NF+3NF, 
chiral 2NF and Bonn-B 2NF, respectively.
Here we represent the channels of the NN system as $^{2S+1}{\cal P}$ 
with the parity ${\cal P}$ and the spin multiplicity $(2S+1)$; 
hence, $^{1}$E, $^{3}$E, $^{1}$O, $^{3}$O channels correspond 
to $(S,T)=(0,1), (1,0), (0,0), (1,1)$ channels, respectively. 
For the triplet ($^{3}$E and $^{3}$O) channels, 
the 2$\pi$-exchange 3NF enhances tensor correlations and hence 
couplings between different states. This makes the imaginary part of 
${\cal U}^{ST}$ more absorptive. 
For the real part, the stronger tensor correlations make 
${\cal U}^{ST}$ more attractive for $^{3}$E 
and less attractive for $^{3}$O. 
For the $^{1}$E channel, the contribution of the 2$\pi$-exchange 3NF 
corresponds to suppressing transitions to 
$\Delta$ resonance due to Pauli blocking, and consequently 
makes ${\cal U}^{ST}$ less attractive.

The ${\cal U}^{ST}$ calculated 
from chiral 2NF (circles) is close to that from Bonn-B 2NF 
(triangles) except for the real part of ${\cal U}^{ST}$ 
in the odd ($^{1}$O and $^{3}$O) channels. 
However, the deviation is not important, 
because the odd components 
hardly contribute to the folding potential for 
NA scattering as mentioned later below Eq. \eqref{G-ST-CE}. 
In general, the $g$-matrix is nonlocal and hence not practical. 
For this reason, 
the $g$-matrix is usually presented by assuming a local form such as 
Gaussian and Yukawa functions. 
The Melbourne group has already constructed a local effective interaction 
on the basis of the nonlocal $g$-matrices calculated 
from Bonn-B 2NF~\cite{Amos}. 
We then introduce the effects of chiral 3NF by multiplying the 
central part $g^{ST}(\vs;k_{\rm F},E)$ of the local Melbourne $g$-matrix 
by the factor 
\bea
f_{}^{ST}(k_{\rm F}^{},E)=
{\cal U}_{({\rm 2NF+3NF})}^{ST}(k_{\rm F}^{},E)
/{\cal U}_{({\rm 2NF})}^{ST}(k_{\rm F}^{},E) , 
\label{fkf}
\eea
where the argument $\vs$ in $g^{ST}(\vs;k_{\rm F},E)$ 
denotes the coordinate between two correlated nucleons and 
${\cal U}_{({\rm 2NF+3NF})}^{ST}(k_{\rm F}^{},E)$ and 
${\cal U}_{({\rm 2NF})}^{ST}(k_{\rm F}^{},E)$ are 
the single-particle potentials with and without chiral-3NF effects, 
respectively. The present prescription is then described by 
\bea
g^{ST}(\vs;k_{\rm F},E) \rightarrow f^{ST}(k_{\rm F},E)
g^{ST}(\vs;k_{\rm F},E) . 
 \label{modified-Melbourne-g-s}
\eea
As mentioned above, the factor $f^{ST}$ is mainly determined from 
the on-shell component of the $g$-matrix. 
However, it should be noted that 
when $V_{123} \neq 0$, the off-shell component of the $g$-matrix 
contributes to the factor $f^{ST}$ through the second term of Eq. \eqref{spp}.

It is shown in Ref. \cite{Koh12} that the spin-orbit part of the $g$-matrix is 
enhanced by chiral 3NF at most by a factor of 4/3. 
This effect is also simply estimated by multiplying the spin-orbit part of 
the Melbourne $g$-matrix by the factor, 
since the effect is small for the present NA scattering.

\subsection{Folding model}
\label{Folding model}

We recapitulate the folding model, following Ref. \cite{Toy13}. 
Since the formalism is parallel between proton and neutron scattering, 
we mainly consider proton scattering as an example. 
Proton elastic scattering can be 
described as a one-body scattering distorted by an optical potential $U$:
\bea
(T_{\viR} + U -E){\Psi}^{(+)}=0 , 
\label{schrodinger}
\eea
where 
$E$ denotes the energy of an incident proton and 
$T_{\viR}$ stands for the kinetic energy with respect to 
the relative coordinate $\vR$ between an incident proton and a target (T). 
The optical potential $U$ can be divided into 
the central (CE), the spin-orbit (LS), and the Coulomb (Coul)
component:
\bea
U=U_{\rm CE}+U_{\rm LS} {\bfi L} \cdot {\bfi \sigma} 
+ V_{\rm Coul} . 
\eea

In the $g$-matrix folding model, $U$ is obtained by folding 
the $g$-matrix with the density of T:
\bea
U(\vR)=\langle \Phi_0 | \sum_{j \in {\rm T}}
g_{pj} | \Phi_0 \rangle \;,
\eea
where $\Phi_0$ is the ground state of T. 
The resulting potential is composed of the direct and exchange parts:
$U=U^{\rm DR}+U^{\rm EX}$. Since the $U^{\rm EX}$ is nonlocal, 
it is localized with 
the Brieva-Rook approximation \cite{Brieva-Rook}. 
The validity of this approximation is shown in 
Ref. \cite{Minomo:2009ds}. 
The central part $U_{\rm CE}$ of the localized $U$ 
is then described as~\cite{Brieva-Rook,CEG,CEG07} 
\bea
U_{\rm CE} \equiv V_{\rm CE}+iW_{\rm CE}
=U_{\rm CE}^{\rm DR}+U_{\rm CE}^{\rm EX}
\eea
with 
\bea
\label{eq:UD}
U_{\rm CE}^{\rm DR}(\vR) \hspace*{-0.15cm} &=& \hspace*{-0.15cm} 
\!\! \sum_{\alpha=p,n} \! \int \!\! \rho_{\alpha}(\vrr) 
            g^{\rm DR}_{p\alpha}(\vs;\rho_{\alpha}) d \vrr, \\
\label{eq:UEX}
U_{\rm CE}^{\rm EX}(\vR) \hspace*{-0.15cm} &=& \hspace*{-0.15cm} - \!\! \sum_{\alpha=p,n} \!
\int \!\! \rho_{\alpha}(\vrr,\vrr-\vs)
            g^{\rm EX}_{p\alpha}(\vs;\rho_{\alpha}) 
            j_0(K s)
            d \vrr ,~~~~
\notag \\
            \label{U-EX}
\eea
where $\vs=\vrr - \vR $ for the coordinate $\vrr$ of an interacting nucleon 
from the center-of-mass (c.m.) of T, and $V_{\rm CE}$ and $W_{\rm CE}$ are 
the real and imaginary parts of $U_{\rm CE}$, respectively. 
Here the mixed density $\rho_{\a}(\vrr,\vR)$ is usually calculated 
with the local Fermi-gas approximation~\cite{Negele}:
\bea
\rho_{{\a}}(\vrr,\vR) \approx 
\rho_{{\a}} \left(\vrr_{m} \right) D_{\a}({\tilde s}) 
\label{LFGA}
\eea
with 
\bea
D_{\a}({\tilde s})=\frac{3}{{\tilde s}^3}
[\sin({\tilde s})-{\tilde s}\cos({\tilde s})], 
\label{LFGA-rm}
\eea
for ${\tilde s}=sk_{\rm F}^{\a}(r_m)$ and the 
midpoint $\vrr_{m}=\vR+\vs/2$ between two correlated nucleons, 
where $k_{\rm F}^{\a}(r_m)$ is related to the local density $\rho_{\a}(r_m)$ 
as $k_{\rm F}^{\a}(r_m)^3= 3\pi^2 \rho_{\a}(r_m)$. 
The local momentum $K(R)$ present in Eq.~\eqref{U-EX} is obtained 
self-consistently, since it is defined as 
$\hbar K(R) \equiv \sqrt{2\mu_{\!R}^{} (E - U_{\rm
CE}-V_{\rm Coul})}$ for the reduced mass $\mu_{\! R}$ of 
the projectile+target system.

The direct and exchange parts of the $g$-matrix interaction, 
$g^{\rm DR}_{p\alpha}$ and $g^{\rm EX}_{p\alpha}$, 
are assumed to be a function 
of the local density $\rho_{\alpha}=\rho_{\alpha}(\vrr_m)$
at the midpoint of the interacting nucleon pair. 
The direct and exchange parts are described by
\bea
&g_{pp}^{\rm DR,EX}(s;\rho_p) = g_{nn}^{\rm DR,EX}(s;\rho_p) 
= \displaystyle\frac{1}{4} \left( \pm g^{01} + 3
 g^{11}\right) ,~~~~\\
 &g_{pn}^{\rm DR,EX} (s;\rho_n) = \displaystyle\frac{1}{8} \left( g^{00} \pm
 g^{01} \pm 3 g^{10} + 3 g^{11}\right) .~~~~
\label{G-ST-CE}
\eea
The even ($^{1}$E and $^{3}$E) components 
of $g^{ST}$ dominate $U_{\rm CE}$, 
since the odd ($^{1}$O and $^{3}$O) components 
are almost canceled each other between $U_{\rm CE}^{\rm DR}$ and 
$U_{\rm CE}^{\rm EX}$. 
%-----LS part-----
The same derivation is possible for the spin-orbit part, 
\bea
U_{\rm LS} \equiv V_{\rm LS}+iW_{\rm LS}
=U_{\rm LS}^{\rm DR}+U_{\rm LS}^{\rm EX} ;
\eea
see Refs. \cite{CEG07,Toy13} for the explicit form of $U_{\rm LS}$.

For heavier targets with the mass number $A$ larger than 40, 
the matter densities are evaluated with 
spherical Hartree-Fock (HF) calculations with 
the Gogny-D1S interaction~\cite{GognyD1S} in which 
the spurious c.m. motions are removed 
in the standard manner~\cite{Sum12}. 
For lighter targets of $A \le 40$, 
the phenomenological proton-density~\cite{phen-density} is taken, 
where the finite-size effect of proton charge 
is unfolded with the standard procedure~\cite{Singhal}. 
The neutron density is assumed to have the same geometry as the 
proton one, since the difference between the neutron root-mean-square radius 
and the proton one is only 1\% in spherical HF calculations.

%Results and Discussions
\section{Results}
\label{Results}

\begin{figure}[tbp]
\begin{center}
 \includegraphics[width=0.4\textwidth,clip]{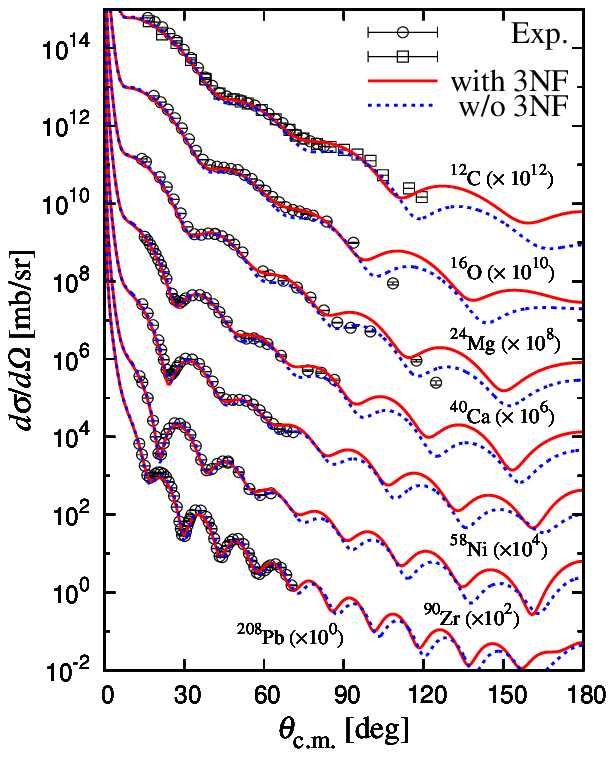}
 \includegraphics[width=0.4\textwidth,clip]{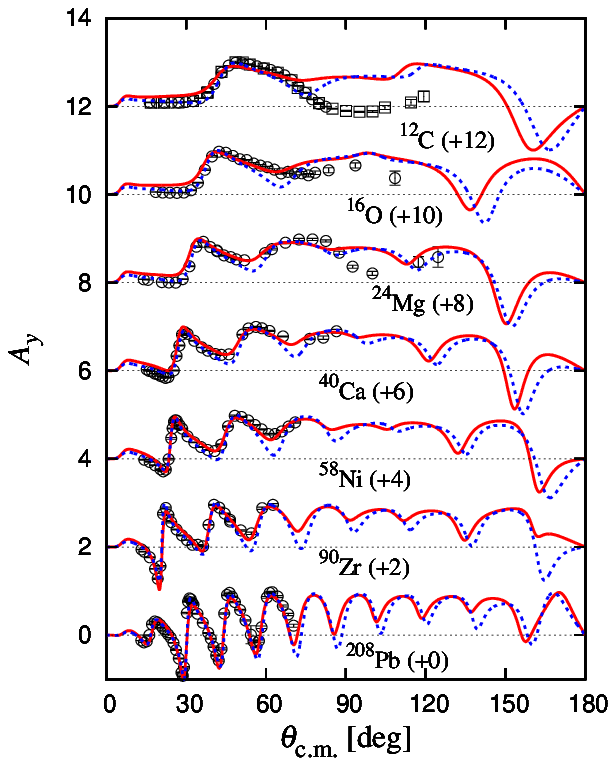}
 \caption{(Color online) 
Differential cross sections and vector analyzing powers for 
proton elastic scattering at 65 MeV. 
The solid (dashed) curves represent the results of 
Melbourne $g$-matrix with (without) chiral-3NF corrections.
Each cross section is multiplied by the factor shown in the figure, 
while each vector analyzing power is shifted up by the number shown 
in the figure. 
Experimental data are taken from Ref. 
\cite{pC12:Ieiri,pC12:Kato,pO16:Mg24:Ca40:Zr90:Sakaguchi,pNi58:Pb208:Sakaguchi}.}
 \label{fig:p+A-scattering}
\end{center}
\end{figure}

\begin{figure}[tbp]
\begin{center}
 \includegraphics[width=0.4\textwidth,clip]{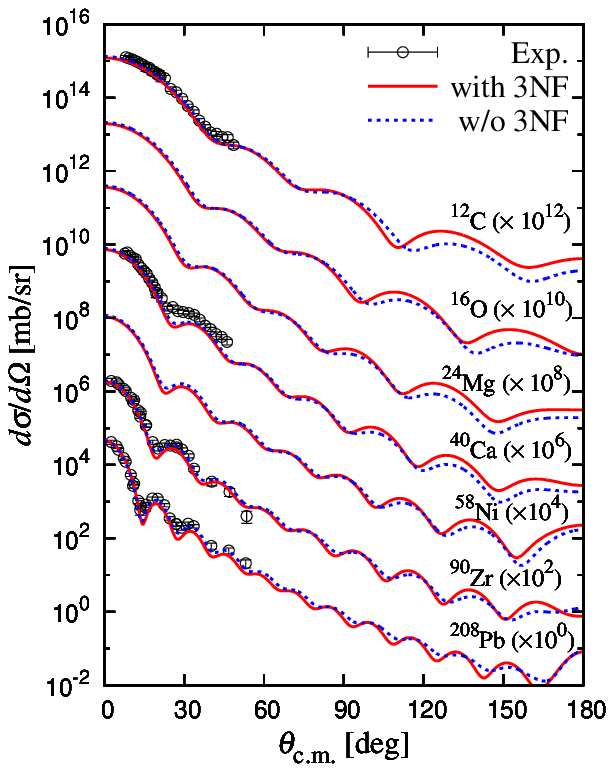}
 \includegraphics[width=0.4\textwidth,clip]{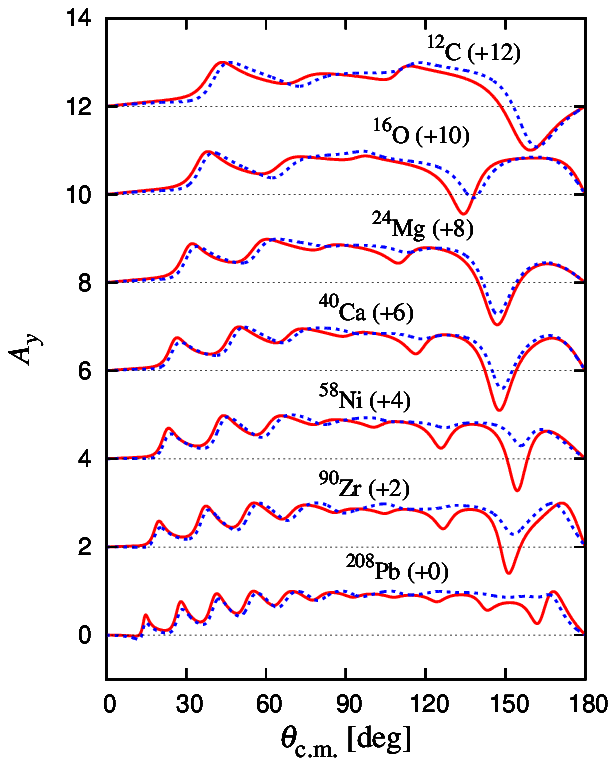}
 \caption{(Color online) 
Differential cross sections and vector analyzing powers for 
neutron elastic scattering at 65 MeV. 
The solid (dashed) curves represent the results of 
Melbourne $g$-matrix with (without) chiral-3NF corrections.
Each cross section is multiplied by the factor shown in the figure, 
while each vector analyzing power is shifted up by the number shown 
in the figure. 
Experimental data are taken from Ref. \cite{nC12:Ca40:Hjort,nZr90:Pb208:Baba}.
}
 \label{fig:n+A-scattering}
\end{center}
\end{figure}

Figure~\ref{fig:p+A-scattering} shows differential cross sections 
$d \sigma/d \Omega$ and vector analyzing powers $A_y$ for 
proton elastic scattering at $E=65$ MeV 
on various  targets from $^{12}$C to $^{208}$Pb. 
The solid (dashed) lines stand for the results of 
Melbourne $g$-matrix with (without) chiral-3NF corrections.
The chiral-3NF effects are small 
at the forward and middle angles 
where the experimental data 
\cite{pC12:Ieiri,pC12:Kato,pO16:Mg24:Ca40:Zr90:Sakaguchi,pNi58:Pb208:Sakaguchi} are available.  
Only an exception is $A_y$ around $\theta = 60^\circ$. 
The chiral-3NF effects improve the agreement with 
the experimental data there. Similar improvement is also seen in 
the previous work 
\cite{Raf13} based on the phenomenological 3NFs. 
This smallness of chiral-3NF effects comes from the fact that 
the effects are significant only in $\rho \ga 0.7 \rho_0$. 
The chiral-3NF effects become significant at backward angles, 
although the experimental data are not available there. 
The backward measurements are thus important 
to investigate chiral-3NF effects. 
Also for neutron scattering at 65 MeV, the chiral-3NF effects are small 
at forward and middle angles and become significant for backward angles, 
as shown in Fig. \ref{fig:n+A-scattering}.

%----------------------
% Figure p+A
%----------------------
%----------------------

%----------------------
% Figure n+A
%----------------------
%----------------------

\begin{figure}[tbp]
\begin{center}
 \includegraphics[width=0.45\textwidth,clip]{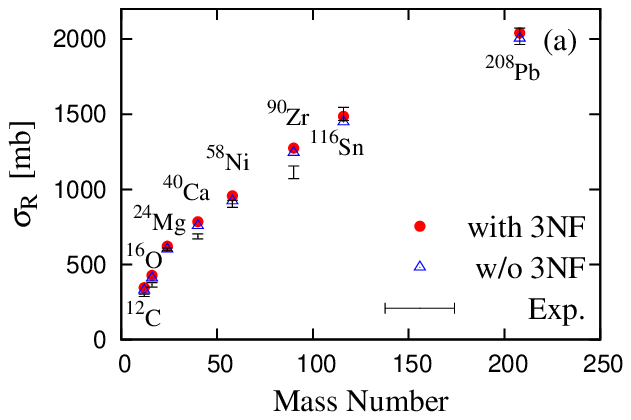}
 \includegraphics[width=0.45\textwidth,clip]{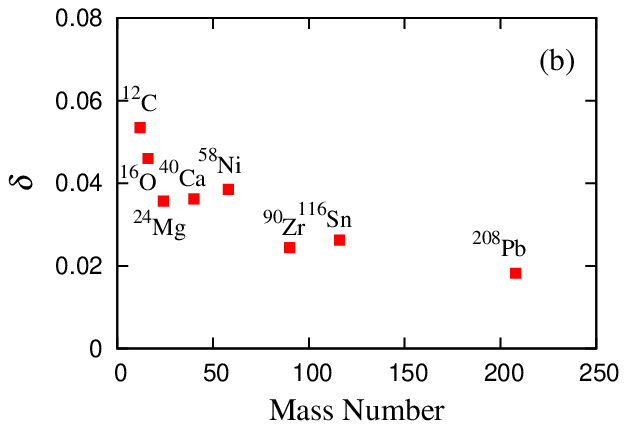}
 \caption{(Color online) 
Reaction cross sections $\sigma_{\rm R}$ 
for proton scattering around $E=65$ MeV. 
The theoretical results are compared with the experimental data 
at $E=65.5$ MeV for $^{12}$C, $^{16}$O, $^{40}$Ca, 
$^{58}$Ni, $^{116}$Sn, $^{208}$Pb, 48.0 MeV for $^{24}$Mg 
and 60.8 MeV for $^{90}$Zr. 
In panel (a) the $\sigma_{\rm R}$ are plotted as a function of $A$. 
The circles (triangles) denote the results of Melbourne $g$-matrix 
with (without) chiral-3NF corrections. 
In panel (b), the relative difference $\delta$ due to chiral 3NF 
is shown as a function of $A$. 
Experimental data are taken from Refs. 
\cite{rcs:Ingemarsson,rcs:Davison,rcs:Menet}.
}
 \label{fig:p+A-scattering:RXS}
\end{center}
\end{figure}

Figure \ref{fig:p+A-scattering:RXS} shows reaction cross sections 
$\sigma_{\rm R}$ for proton scattering around $E$=65 MeV on 
various  targets from $^{12}$C to $^{208}$Pb. 
In panel (a), the $\sigma_{\rm R}$ are plotted as a function of target mass 
number $A$.
The circles (triangles) stand for the results of Melbourne $g$-matrix 
with (without) chiral-3NF corrections. 
The effects of chiral 3NF on $\sigma_{\rm R}$ are small, so that 
both the Melbourne $g$-matrix and the modified Melbourne $g$-matrix 
with chiral-3NF corrections reproduce the measured $\sigma_{\rm R}$. 
The agreement of the theoretical results with the experimental data are 
particularly good for heavier targets such as $^{116}$Sn and $^{208}$Pb 
where the local density approximation is considered to be good. 
In panel (b), the relative difference 
\bea
\delta=\frac{\sigma_{\rm R}^{\rm 2NF+3NF}-\sigma_{\rm R}^{\rm 2NF}}
{\sigma_{\rm R}^{\rm 2NF}}
\eea
is plotted as a function of $A$. The chiral 3NF enhances $\sigma_{\rm R}$ 
only by a few percent, so that the Melbourne $g$-matrix keeps good agreement 
with the experimental data.

%----------------------
% Figure p+A; reaction cross section
%----------------------
%----------------------

\begin{figure}[htbp]
\begin{center}
 \includegraphics[width=0.5\textwidth,clip]{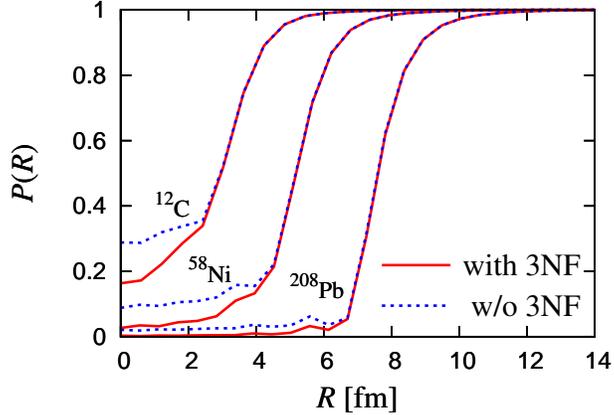}
 \caption{(Color online) The probability $P(R)$ of elastic scattering 
as a function of $R$ for proton scattering from $^{12}$C, $^{58}$Ni and 
$^{208}$Pb at $E=65$~MeV. The solid (dashed) lines represent the results 
of Melbourne $g$-matrix with (without) chiral-3NF corrections. 
   }
\label{Fig:S-mat;p+Ni;E=65}
\end{center}
\end{figure}

\begin{figure}[htbp]
\begin{center}
 \includegraphics[width=0.45\textwidth,clip]{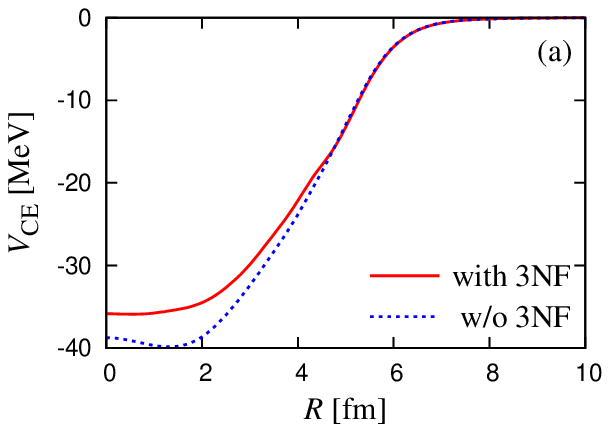}
 \includegraphics[width=0.45\textwidth,clip]{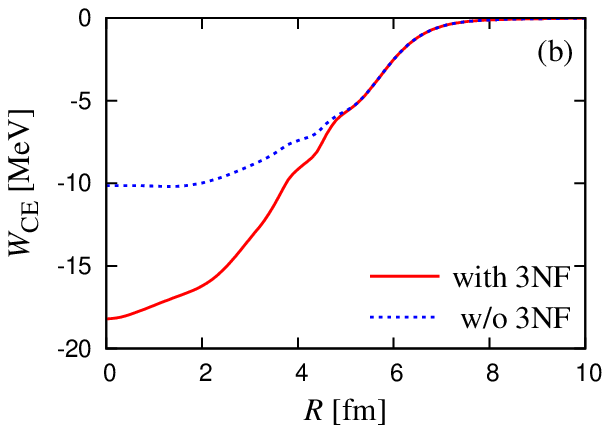}
 \caption{(Color online) 
$R$ dependence of the central part of the folding potential 
for $p$+$^{58}$Ni elastic scattering at $E=65$~MeV. 
Panels (a) and (b) correspond to the real and imaginary parts, respectively. 
The solid (dashed) lines represent the results 
of Melbourne $g$-matrix with (without) chiral-3NF corrections. 
}
\label{Fig:pot;p+Ni;E=65}
\end{center}
\end{figure}

The probability $P(R)$ of elastic scattering at each $R$ can be described by 
the elastic $S$-matrix element $S_L$ as $P(R)=|S_L|^2$, where 
$R$ can be estimated from the relative angular momentum $L$ 
between proton and T with the semi-classical relation $L=R K(\infty)$. 
Figure \ref{Fig:S-mat;p+Ni;E=65} shows $P(R)$ as a function of $R$ 
for proton scattering from $^{12}$C, $^{58}$Ni and $^{208}$Pb at $E=65$~MeV. 
The solid (dashed) lines stand for the results of Melbourne $g$-matrix 
with (without) chiral-3NF corrections. 
The chiral-3NF effects appear mainly in the inner region of T 
where $P(R)$ is small, and make $P(R)$ even smaller. 
In Fig. \ref{Fig:pot;p+Ni;E=65}, the central part of $U$ is plotted 
as a function of $R$ for $^{58}$Ni. 
The chiral 3NF makes $U$ more absorptive and 
less attractive. The effects in the peripheral region of $R \ga 4$~fm 
affect $d \sigma/d \Omega$ and $A_y$ at the forward and middle angles 
where the experimental data are available. 
In the inner region of $R \la 4$~fm, 
the chiral 3NF little affects $d \sigma/d \Omega$ and $A_y$ 
at the forward and middle angles, 
since $P(R)$ is already small in the results of Melbourne $g$-matrix 
without chiral-3NF corrections.

In the folding procedure, the even ($^{1}$E and $^{3}$E) components 
of $g^{ST}$ dominate $U_{\rm CE}$, 
since the odd ($^{1}$O and $^{3}$O) components 
are almost canceled each other between $U_{\rm CE}^{\rm DR}$ and 
$U_{\rm CE}^{\rm EX}$. 
This means that the repulsive effect of chiral 3NF on $V_{\rm CE}$ comes 
from the same effect on ${\cal U}^{ST}$ in the $^{1}$E channel, that is, 
from the suppression of transitions to $\Delta$ resonance due to Pauli blocking. 
Similarly, the strong absorption effect of chiral 3NF 
on $W_{\rm CE}$ is originated in the enhancement of tensor correlations 
due to chiral 3NF.

%%%%%%%%%%%%%%%%%%%%%%%
%%%  Figure; S-matrix
%%%%%%%%%%%%%%%%%%%%%%%

%%%%%%%%%%%%%%%%%%%%%%%
%%%  Figure; potential
%%%%%%%%%%%%%%%%%%%%%%%

%Summary
\section{Summary} 
\label{Summary} 

We have investigated the roles of chiral NNLO 3NF 
in NA elastic scattering, using the standard framework based on the BHF method 
for nuclear matter and the folding model for NA scattering. 
Ch-EFT is a definite way of organizing interactions among 
nucleons, and consequently, 2NF and 3NF are consistently defined. 
The present framework based on Ch-EFT has been applied to NA scattering 
at a lower incident energy  of $E=65$ MeV over various targets, 
since Ch-EFT is more reliable for lower $E$.

BHF calculations were done for positive energy
with chiral N$^{3}$LO 2NF including NNLO 3NF with the cutoff $\Lambda$ 
of 550 MeV on the basis that the same calculations 
for negative energies well reproduce empirical 
saturation properties of symmetric nuclear matter. 
The single-particle potential calculated from chiral 2NF+3NF deviates 
from that from chiral 2NF in the density region $\rho \ga 0.7 \rho_0$. 
The difference mainly comes from the 2$\pi$-exchange diagram. 
The diagram generates absorptive corrections in the triplet channels by 
enhancing tensor correlations and repulsive corrections 
in the singlet $^{1}$E channel by suppressing transitions to $\Delta$ 
resonance due to Pauli blocking. 
The repulsive contribution in the $^{1}$E channel dominates the effects of 
chiral 3NF on the real part of the $g$-matrix.

The effects of chiral 3NF are incorporated in the folding potential with 
the following simple procedure, 
as the first estimate of chiral-3NF effects on NA scattering. 
The Melbourne group has already constructed the local effective interaction 
on the basis of the $g$-matrices from Bonn-B 2NF. 
The single-particle potential calculated from chiral 2NF 
with $\Lambda=550$ MeV is found to be close to that from Bonn-B 2NF. 
We have then modified the Melbourne $g$-matrix so as to 
reproduce the single-particle potentials obtained from chiral 2NF+3NF. 
In the procedure, the effects of chiral 3NF on the on-shell 
component of the $g$-matrix are approximately taken into account. 
The chiral-3NF effects are small for differential cross sections 
and vector analyzing powers at the forward and middle angles 
where the experimental data are available, 
but the effects surely improve the agreement with 
measured vector analyzing powers around middle angles. 
Similar improvement is also seen in the previous work \cite{Raf13} 
based on phenomenological 3NFs.

Chiral 3NF, mainly in its the 2$\pi$-exchange diagram, 
makes the folding potential less attractive and more absorptive. 
In the previous work \cite{Raf13}, phenomenological 3NFs make the potential 
less attractive but less absorptive. Thus, chiral 3NF yields 
a different property for the imaginary part of the folding potential. 
This novel property is originated in the enhancement of tensor correlations 
due to the 2$\pi$-exchange diagram. 
Ch-EFT, furthermore, says that the repulsive effect of the diagram on 
the folding model comes from the suppression of transitions 
to $\Delta$ resonance due to Pauli blocking.

Owing to the density dependence of 3NF contributions, the chiral-3NF effects 
are sizable in the inner region of target, but small 
in the peripheral region. The large effect in the inner region 
is, however, masked by the strong absorption of the incident flux. 
Consequently, the chiral-3NF effects are small for NA scattering 
at the forward and middle angles where the experimental data are 
available at present. 
This is the reason why the Melbourne $g$-matrix with no 3NF effects 
well accounts for measured cross sections and vector analyzing powers 
for NA scattering. 
If the measurements are made at backward angles, 
the data should reveal chiral-3NF effects. 
Another possibility of detecting the chiral-3NF effects is in 
transfer reactions such as $(d, p)$ reactions, 
since the optical potential itself changes sizably with the effects.   
Furthermore, AA scattering is also interesting, since 
the density higher than $0.7\rho_{0}$ is certainly realized in the scattering. 
We discuss this subject in a separate paper \cite{Minomo:2014eqa}
by using the double-folding model.

%Acknowledgement
\section*{Acknowledgements} 
This work is supported in part by
by Grant-in-Aid for Scientific Research
(Nos. 244137, 25400266, and 26400278)
from Japan Society for the Promotion of Science (JSPS).

%%--------------------------------------------------------------------%%
%%                           References                               %%
%%--------------------------------------------------------------------%%

\end{document}